\documentclass[twocolumn,10pt]{article}

\usepackage[margin=1in]{geometry}
\usepackage{graphicx}
\usepackage{amsmath,amssymb}
\usepackage{booktabs}
\usepackage[utf8]{inputenc}
\usepackage[T1]{fontenc}
\usepackage{listings}
\usepackage{xcolor}
\usepackage{enumitem}
\usepackage[numbers,sort&compress]{natbib}
\usepackage{fancyhdr}
\usepackage{authblk}
\usepackage{float}
\usepackage{flafter}
\usepackage{microtype}
\usepackage{ragged2e}
\usepackage[font=small,labelfont=bf]{caption}
\usepackage[hidelinks]{hyperref}

\definecolor{codegray}{rgb}{0.5,0.5,0.5}

\newcommand{\poduid}{\texttt{pod\_\allowbreak uid}}
\newcommand{\podspechash}{\texttt{pod\_\allowbreak spec\_\allowbreak hash}}
\newcommand{\workloadid}{\texttt{workload\_\allowbreak id}}
\newcommand{\reportdata}{\texttt{report\_\allowbreak data}}
\newcommand{\spechash}{\texttt{spec\_\allowbreak hash}}
\newcommand{\securitypolicydigest}{\texttt{security\_\allowbreak policy\_\allowbreak digest}}
\newcommand{\securitypolicyschemaversion}{\texttt{security\_\allowbreak policy\_\allowbreak schema\_\allowbreak version}}
\newcommand{\wgpubkey}{\texttt{wg\_\allowbreak pubkey}}
\newcommand{\wgendpoint}{\texttt{wg\_\allowbreak endpoint}}

\title{\textbf{Implement Kubernetes Pod-Level Remote Attestation \\ for Confidential Workloads on dstack}}

\author{
Yang Yang, Kevin Wang$^\dagger$, Yuanhai Luo, Hang Yin$^\dagger$, Jie Cai, Shunfan Zhou$^\dagger$, Wenfeng Wang$^\dagger$\\
OPPO, $^\dagger$Phala\\
\texttt{\{yangyang1, null.luo, Gload\}@oppo.com}\\
$^\dagger$\texttt{\{kvinwang, hangyin, shelvenzhou, wfwang\}@phala.network}
}

\date{}

\begin{document}

\maketitle

\begin{abstract}
The rise of LLM-as-a-Service and other confidential cloud workloads demands cryptographic proof that user data is processed in a trusted, untampered environment. Existing solutions, notably Confidential Containers (CoCo), enforce a strict ``one Pod per VM'' model that attests only the Guest OS stack, leaving container-level identity unverified and incurring prohibitive per-VM resource overhead. We present \textbf{dstack-capsule}, a Kubernetes platform that enables \textbf{Pod-level remote attestation} on Intel TDX by allowing multiple Pods to share a single Confidential VM while each retains independent, hardware-backed proof of identity. Our key insight is a two-layer attestation architecture: static platform measurements are frozen in RTMR[3] via an irreversible \textit{privilege fuse}, while dynamic Pod identities (\poduid, \podspechash, \workloadid) are embedded in the TDX Quote's \reportdata{} field and signed by hardware on every request. dstack-capsule introduces (1) a Pod-level attestation protocol binding Pod spec digests to hardware-signed Quotes; (2) a privilege fuse mechanism that atomically transitions a node from setup mode to secure mode; (3) a multi-layer sandbox spanning storage, runtime, admission, API, and network isolation layers; and (4) a complete open-source implementation based on Kubernetes 1.32, Intel TDX, and Sysbox. We evaluate the security properties, attestation correctness, and performance characteristics of dstack-capsule, demonstrating that it achieves Pod-granularity verification without the resource overhead of per-VM isolation.
\end{abstract}

\section{Introduction}

\subsection{Motivation}

Cloud-native confidential computing has become essential for AI inference, financial analytics, and healthcare workloads handling sensitive data. In the LLM-as-a-Service paradigm, users submit proprietary prompts to cloud providers expecting their data to remain confidential during processing---a guarantee that conventional cloud infrastructure cannot provide. Regulatory frameworks increasingly mandate cryptographic protection for data in use, complementing existing encryption for data at rest and in transit.

Trusted Execution Environments (TEEs), particularly Intel Trust Domain Extensions (TDX), provide hardware-isolated execution contexts with built-in remote attestation capabilities~\citep{intel2023tdx}. These enable a fundamental guarantee: a remote verifier can cryptographically confirm that a specific, untampered software stack executes on genuine hardware. However, bridging this hardware primitive to the Kubernetes abstraction layer---where workloads are dynamically scheduled, scaled, and migrated---introduces architectural challenges that existing solutions have not fully resolved~\citep{kubernetes2025docs}.

\subsection{Problem Statement}

Confidential Containers (CoCo) represents the current state-of-the-art for deploying confidential workloads on Kubernetes~\citep{coco2023confidential}. CoCo integrates Kata Containers with TEEs to launch each Pod inside its own lightweight VM~\citep{katacontainers2018}, with an attestation agent running within the VM's Guest OS. While providing strong isolation, this design suffers from three fundamental challenges:

\noindent\textbf{Challenge 1: VM-level attestation granularity.} CoCo attests the Guest OS stack (kernel, initrd, rootfs) but does not measure container image content, Pod specifications, or workload identity. A verifier can confirm that ``a trusted Linux kernel is running'' but cannot cryptographically verify that ``this specific Pod with these exact container images and environment variables is executing.''

\noindent\textbf{Challenge 2: Resource overhead.} The ``one Pod = one VM'' model means each Pod carries the full memory and CPU cost of a guest kernel, VM management layer, and attestation agent. For high-density Kubernetes workloads, this overhead becomes prohibitive: a 64\,GB host can accommodate approximately 30 CoCo Pods before memory exhaustion, substantially fewer than what the same hardware can support for non-TEE workloads.

\noindent\textbf{Challenge 3: No Pod-level network isolation proof.} CoCo provides no mechanism to cryptographically prove that a specific Pod's network connectivity is restricted to a declared whitelist.

These challenges motivate our core research question: \textbf{How can we provide independent, hardware-backed, Pod-granularity remote attestation while allowing multiple Pods to efficiently share a single TEE resource?}

\subsection{Key Insight and Contributions}

Our key insight is that TEE hardware measurement registers (RTMR) are append-only and immutable, making them fundamentally unsuitable for measuring dynamic, short-lived Kubernetes Pods. Instead, we decouple attestation into two layers:

\begin{itemize}[nosep]
\item \textbf{Static platform attestation (RTMR[3]):} Proves that the CVM runs an expected, unmodified OS and system software stack. This is measured once at boot and frozen by an irreversible \textit{privilege fuse}.
\item \textbf{Dynamic Pod attestation (\reportdata):} Each Pod's identity (\poduid, \podspechash, \workloadid) is embedded in the TDX Quote's 64-byte \reportdata{} field, which the CPU signs on every attestation request.
\end{itemize}

This two-layer design enables multiple Pods to share a single CVM while each retains independent, hardware-backed cryptographic proof of identity. We make four contributions:

\begin{enumerate}[nosep]
\item \textbf{Pod-level remote attestation protocol.} We present the first system that binds Pod-specific identifiers to hardware-signed TDX Quotes via the \reportdata{} field, with verification covering Pod spec digests, caller identity via UDS peer credentials combined with cgroup parsing, and four independent validation steps.

\item \textbf{Privilege fuse mechanism.} We design an irreversible, atomic state transition (compare-and-swap plus persistent marker file) that moves a Kubernetes node from a privileged setup phase to a secure runtime phase, with RTMR[3] frozen at the transition point and all subsequent privileged Pods rejected.

\item \textbf{Multi-layer sandbox.} We integrate five isolation layers---storage (ZFS per-Pod encryption), runtime (Sysbox user namespaces), admission (privilege rejection), API (default-deny with opt-in), and network (WireGuard VPC with whitelists)---to contain the blast radius of Pod compromise within a shared CVM.

\item \textbf{Complete implementation and evaluation.} We implement dstack-capsule based on Kubernetes 1.32, Intel TDX, and Sysbox ($\sim$7,700 LoC in Rust across core binaries, $\sim$660 LoC in Go), and evaluate its security properties, attestation correctness, and performance against the CoCo baseline.
\end{enumerate}

\section{Background and Threat Model}

\begin{table*}[ht]
\centering
\begin{tabular}{ll}
\toprule
\textbf{Threat Capability} & \textbf{Defense Mechanism} \\
\midrule
Hypervisor compromise & TEE hardware isolation (TDX memory encryption) \\
OS image tampering & dm-verity + MRTD/RTMR[0--2] verification \\
Malicious agent injection & RTMR[3] measures agent config + Ed25519 key \\
Malicious Pod scheduling & KMS workload authorization + spec digest check \\
Container escape & Sysbox user namespace (UID 0 $\rightarrow$ 65536) \\
Host network access & Privilege fuse: deny hostNetwork/hostPath \\
Cross-Pod data access & Per-Pod ZFS AES-256-GCM encryption \\
Attestation forgery & UDS peer credentials + cgroup parsing \\
Privilege re-enablement & Irreversible fuse (CAS + persistent marker) \\
\bottomrule
\end{tabular}
\caption{Threat model: capabilities and corresponding defense mechanisms.}
\label{tab:threats}
\end{table*}

\subsection{Intel TDX Primer}

Intel Trust Domain Extensions (TDX) provides hardware-based isolation for virtual machines called Trust Domains (TDs)~\citep{intel2023tdx}. Each TD executes in a CPU-protected memory region inaccessible to the hypervisor, the host OS, or other TDs.

\textbf{Measurement Registers.} TDX maintains two types of measurement registers:

\begin{itemize}[nosep]
\item \textbf{MRTD (Measurement Register of Trust Domain):} A 48-byte SHA-384 hash computed by hardware during TD creation, measuring the OVMF firmware. MRTD is immutable for the TD's lifetime.
\item \textbf{RTMR[0--3] (Runtime Measurement Registers):} Four 48-byte append-only registers. Each extend operation computes $\text{RTMR}_\text{new} = \text{SHA-384}(\text{RTMR}_\text{old} \parallel \text{data})$. Once extended, measurements cannot be removed or modified.
\end{itemize}

\textbf{Quote Generation.} A TDX Quote is a cryptographically signed attestation report generated by the CPU. It contains MRTD, all four RTMR values, a 64-byte \reportdata{} field (caller-provided), and a hardware signature verifiable via Intel's Data Center Attestation Primitives (DCAP)~\citep{intel2020dcap}.

\textbf{dm-verity and UKI.} The root filesystem is protected by dm-verity (device-mapper integrity), which maintains a Merkle hash tree over all filesystem blocks. The root hash is embedded in the UKI (Unified Kernel Image)---a single EFI executable combining the kernel, initrd, and boot parameters---which is measured into RTMR[1] during boot.

\subsection{Confidential Containers and Design Challenges}

Confidential Containers (CoCo) is a CNCF sandbox project that enables TEE-based confidential computing for Kubernetes~\citep{coco2023confidential}. Its architecture uses Kata Containers to launch each Pod inside a dedicated lightweight VM~\citep{katacontainers2018}. An attestation agent runs within each VM's Guest OS and generates TEE attestation reports.

\textbf{Attestation Granularity.} CoCo attests the Guest OS stack but does not measure container image content. The rationale is practical: container images can be large, and measuring them into RTMR would be prohibitively slow. However, this means a verifier can only confirm that ``a trusted OS is running''---not ``the expected application container is running.''

\textbf{Resource Model.} The ``one Pod = one VM'' design creates significant overhead. Each VM requires dedicated memory for the guest kernel, user-space runtime, and attestation agent, approximately 2\,GB per Pod.

\textbf{Network Isolation.} CoCo does not provide any mechanism to cryptographically prove network isolation policies at Pod granularity.

\subsection{Threat Model}

\textbf{Adversary.} We consider a powerful adversary comprising: (1) the cloud platform operator, who controls the hypervisor, host OS, and Kubernetes control plane; and (2) the Pod developer, who provides the container image and may attempt to extract user data or escalate privileges. We assume the adversary may control both entities simultaneously (collusion).

\textbf{Trust Assumptions.} Our security guarantees rely on: Intel TDX hardware and microcode; the open-source, auditable OS image protected by dm-verity; and the KMS running in an independent TEE.

\textbf{Out of Scope.} Side-channel attacks, application-layer protocol vulnerabilities, hardware microcode bugs, or denial-of-service attacks.

\textbf{Data Leakage Vectors and Defenses.} Table~\ref{tab:threats} maps each threat capability to its corresponding defense mechanism.

\section{System Design}

\subsection{Architecture Overview}

\begin{figure*}[tp]
\centering
\includegraphics[width=0.82\textwidth]{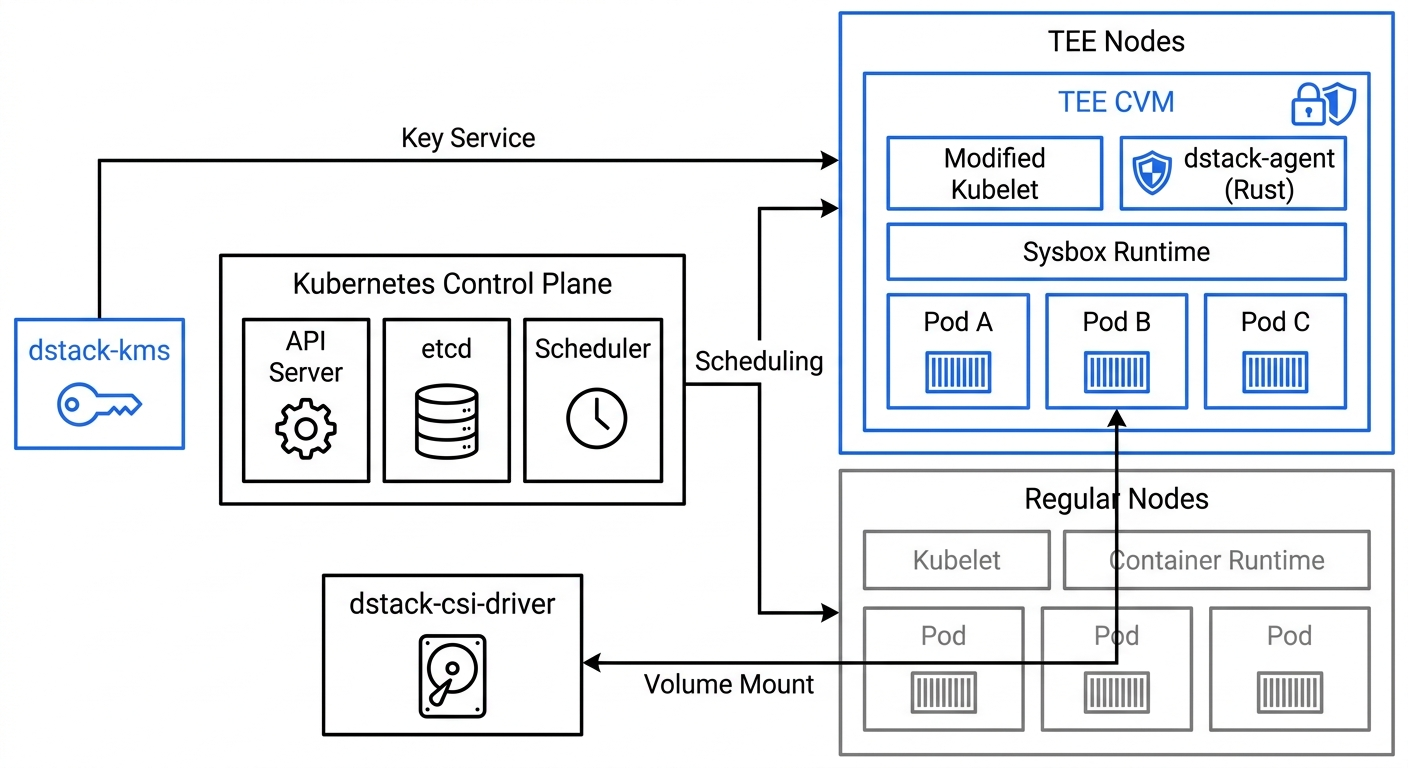}
\caption{dstack-capsule hybrid cluster architecture.}
\label{fig:arch}
\end{figure*}

Figure~\ref{fig:arch} depicts the hybrid cluster architecture. dstack-capsule deploys TEE-enabled nodes alongside regular Kubernetes nodes under a shared control plane. The system comprises four core components:

\begin{itemize}[nosep]
\item \textbf{dstack-agent:} Runs on each TEE node. It mediates Pod admission, computes Pod spec digests, generates TDX Quotes with embedded \reportdata, and serves as the Authorizer gRPC backend.
\item \textbf{dstack-kms:} Runs in a separate TEE and provides key management through RA-TLS mutual attestation.
\item \textbf{dstack-csi-driver:} Creates per-Pod encrypted ZFS datasets and binds storage keys to attested workload identity.
\item \textbf{Modified Kubelet:} Adds Authorizer Hook points for Pod admission, lifecycle events, measurement queries, and Kubelet API authorization.
\end{itemize}

\subsection{Two-Phase Node Lifecycle}

\begin{figure}[tbp]
\centering
\includegraphics[width=0.88\columnwidth]{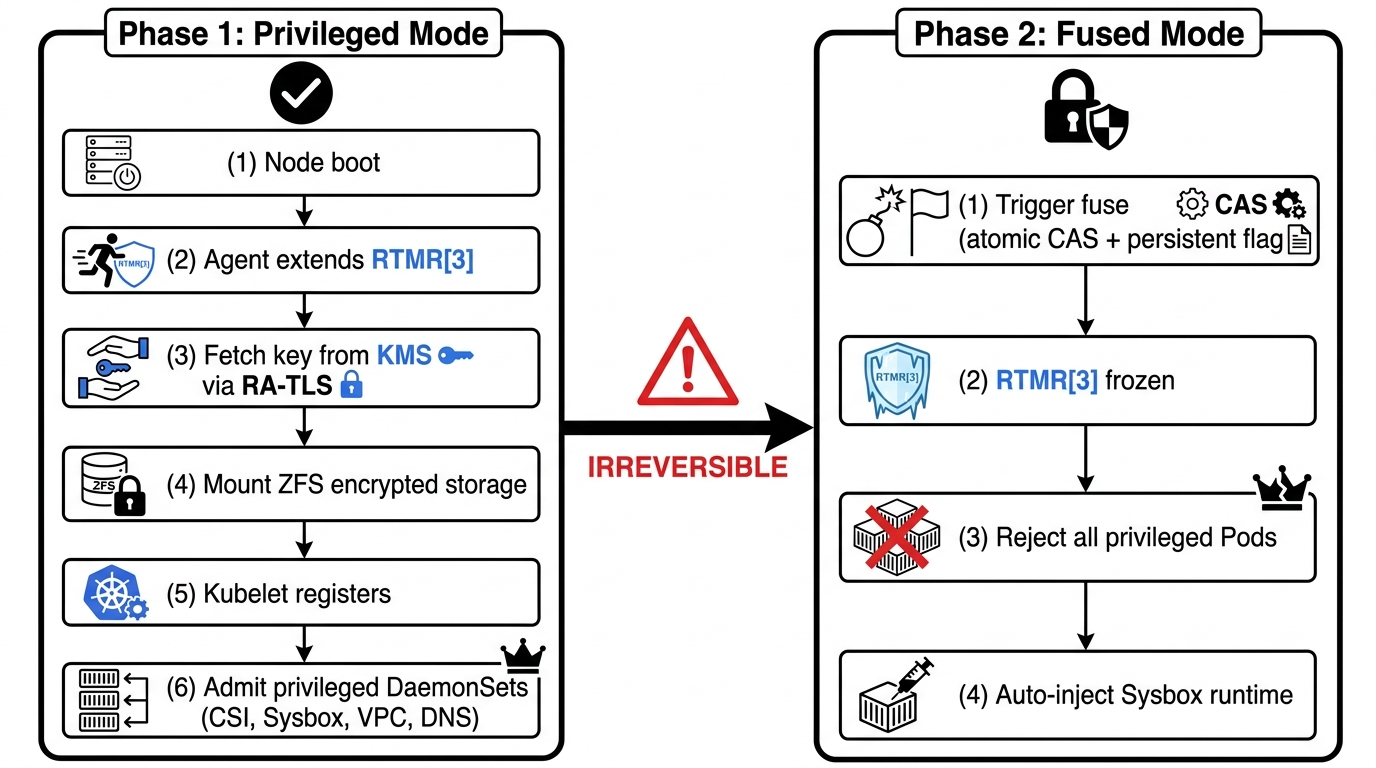}
\caption{Two-phase node lifecycle state machine. Phase 1 (Privileged Mode) allows privileged DaemonSets; Phase 2 (Fused Mode) irreversibly freezes RTMR[3] and rejects all privileged Pods.}
\label{fig:lifecycle}
\end{figure}

Figure~\ref{fig:lifecycle} illustrates the state machine. The node lifecycle splits into two irreversibly separated phases:

\textbf{Phase 1: Privileged Mode.} When a node first boots:
\begin{enumerate}[nosep]
\item dstack-agent starts and extends RTMR[3] with its configuration, node ID, system workload ID, and Ed25519 public key.
\item Agent requests node-level encryption keys from KMS via RA-TLS (Remote Attestation TLS, a mutual attestation protocol where both parties prove TEE identity).
\item Agent mounts ZFS encrypted storage.
\item Kubelet starts and registers with the control plane.
\item The node admits privileged DaemonSets (CSI, Sysbox, VPC, DNS), each with a matching \texttt{workload-id} annotation.
\end{enumerate}

\textbf{Phase 2: Fused Mode.} Once system components are deployed:
\begin{enumerate}[nosep]
\item The cluster controller triggers the privilege fuse.
\item Agent performs an atomic CAS (compare-and-swap) on an internal flag and writes a persistent marker file. This transition is \textbf{irreversible}.
\item RTMR[3] is frozen---no further extensions occur.
\item The node rejects all new Pods requesting privileged capabilities (hostNetwork, hostPID, hostIPC, hostPath, privileged).
\item All new Pods are automatically injected with Sysbox runtime and the Pod API Unix Domain Socket (UDS).
\end{enumerate}

\subsection{Pod-Level Remote Attestation}

\subsubsection{Why Pods Are Not Measured into RTMR}

A na\"ive approach would extend RTMR[3] with each Pod's spec digest. We reject this for three reasons: (1) Pod lifecycle is dynamic, constantly changing RTMR values; (2) verifiers would need complete Pod admission history; (3) RTMR accumulation causes unbounded log growth and violates the append-only invariant's intended semantics.

\subsubsection{Two-Layer Attestation Design}

\begin{figure}[tbp]
\centering
\includegraphics[width=\columnwidth]{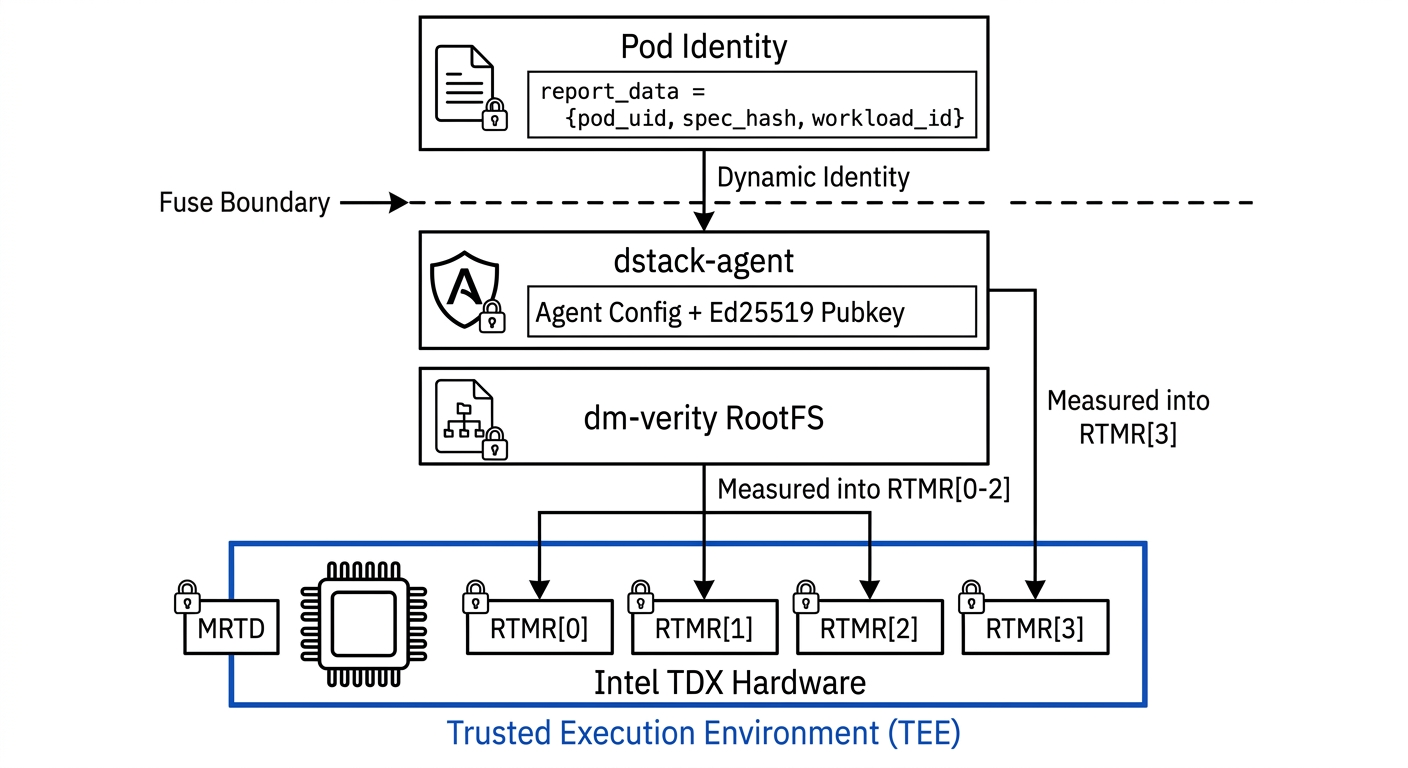}
\caption{TDX trust chain --- MRTD/RTMR[0--2] measure static platform; RTMR[3] measures dstack-agent and freezes at the Fuse Boundary; \reportdata{} carries dynamic Pod identity above.}
\label{fig:trust}
\end{figure}

Figure~\ref{fig:trust} shows the trust chain. Attestation operates at two independent layers:

\begin{itemize}[nosep]
\item \textbf{Layer 1 (Platform Trust):} MRTD and RTMR[0--3] prove the CVM runs an expected OS and agent. The agent's Ed25519 public key is recorded in RTMR[3], binding all subsequent agent signatures to the hardware measurement.
\item \textbf{Layer 2 (Pod Identity):} \reportdata{} contains SHA-256 of \texttt{\{user\_\allowbreak report\_\allowbreak data, pod\_\allowbreak uid, pod\_\allowbreak spec\_\allowbreak hash, workload\_\allowbreak id\}}. This 64-byte field is signed by the CPU alongside the RTMR values, creating a hardware-backed binding between Pod identity and platform state.
\end{itemize}

\subsubsection{Pod Spec Digest}

\begin{enumerate}[nosep]
\item \textbf{Canonicalization:} Pod spec digests use JCS (JSON Canonicalization Scheme, RFC 8785) via an embedded QuickJS engine, which produces a deterministic byte representation independent of JSON key ordering or whitespace~\citep{rundgren2020jcs}. The \reportdata{} payload construction uses \texttt{serde\_json\_canonicalizer}, a pure-Rust JCS implementation, for the attestation path.
\item \textbf{Hashing:} SHA-384 produces a 48-byte digest covering all Pod fields (containers, images, environment variables, volumes, commands, resource limits).
\item \textbf{Customization:} Custom canonicalization scripts are supported via the \texttt{dstack.org/\allowbreak canonicalize-script} annotation for workloads with specialized digest requirements.
\end{enumerate}

\textbf{Determinism Guarantee.} JCS (RFC 8785) guarantees deterministic output for any valid JSON input, independent of parser-specific serialization behavior. The SHA-384 hash over the JCS output therefore produces a consistent digest regardless of how the Pod spec was originally formatted. We verify this property in our test suite by canonicalizing the same Pod spec through multiple JSON encoders and confirming identical digests.

\subsubsection{Pod Proof}

An Ed25519 signature over the colon-separated concatenation of 11 fields:
\begin{lstlisting}
namespace:workload_id:pod_uid:digest:security_policy_digest:
  security_policy_schema_version:wg_pubkey:wg_endpoint:
  node_id:issued_at:expires_at
\end{lstlisting}
The signing key's public component is bound via RTMR[3], and KMS verifies this signature before returning any workload-specific keys. The \securitypolicydigest{} and \securitypolicyschemaversion{} fields bind the Pod's network isolation policy to the proof; \wgpubkey{} and \wgendpoint{} support WireGuard VPC network isolation verification.

\subsubsection{Attestation Generation}

\begin{figure*}[tp]
\centering
\includegraphics[width=0.78\textwidth]{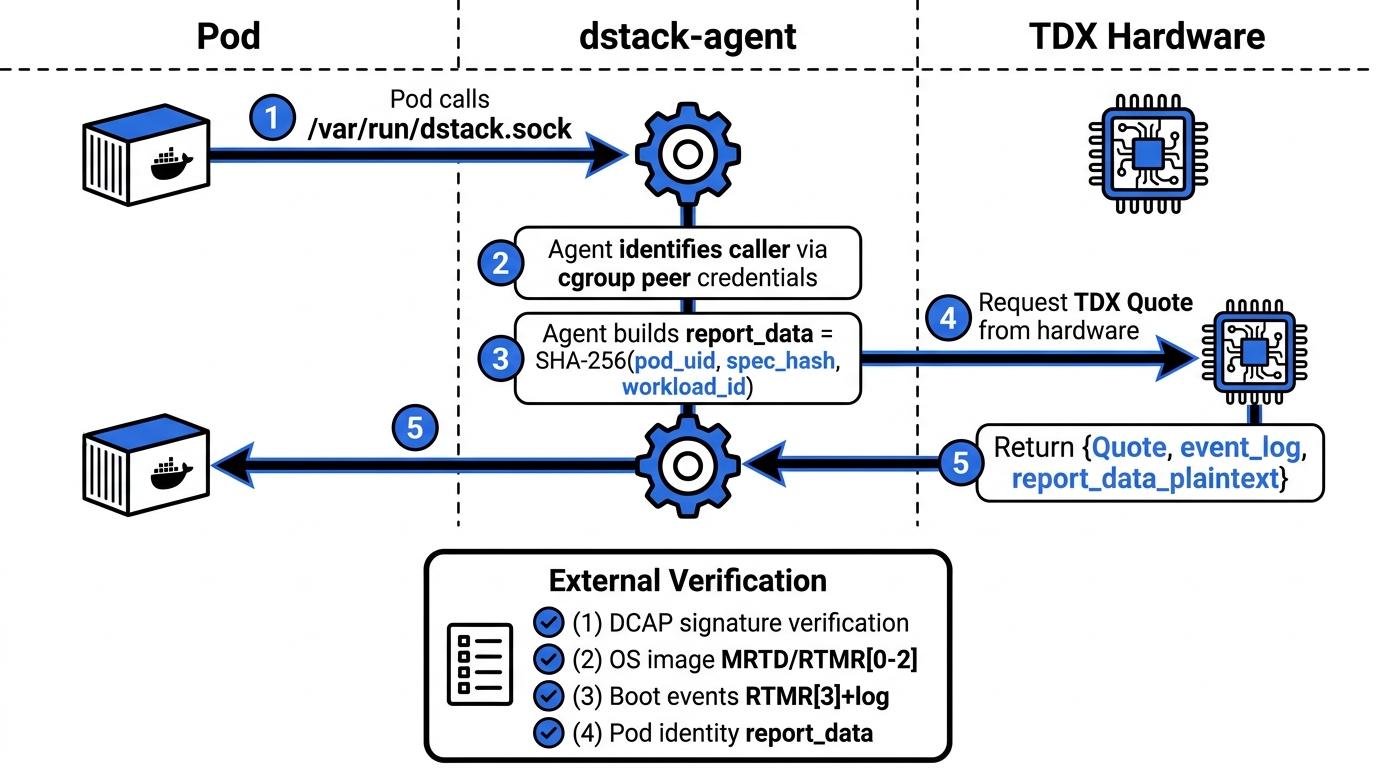}
\caption{Pod-level attestation flow: Pod $\rightarrow$ Agent $\rightarrow$ TDX, with SO\_PEERCRED caller binding and four-stage verification.}
\label{fig:attest}
\end{figure*}

Figure~\ref{fig:attest} shows the sequence. The attestation flow proceeds as follows:
\begin{enumerate}[nosep]
\item Pod calls \texttt{Attest} API via read-only UDS (Unix Domain Socket) at \texttt{/var/run/dstack.sock}.
\item Agent identifies caller via UDS peer credentials (\texttt{SO\_PEERCRED}) combined with cgroup parsing (\texttt{/proc/\{pid\}/cgroup})---a two-step kernel-enforced mechanism that returns the actual process identity, unforgeable even by root inside the container.
\item Agent constructs \reportdata{} and requests TDX Quote from hardware.
\item Returns the TDX Quote, associated event log, and plaintext \reportdata{} to the caller.
\end{enumerate}

\subsubsection{External Verification}

Verification requires four independent checks: (1) Quote signature via Intel DCAP; (2) OS image via MRTD/RTMR[0--2] against a whitelist; (3) Node boot events via RTMR[3] + event log; (4) Pod identity via \reportdata{} SHA-256 match against independently computed \podspechash.

\subsection{KMS and Key Derivation}

Keys are derived using HKDF (HMAC-based Extract-and-Expand Key Derivation Function, RFC 5869)~\citep{krawczyk2010hkdf}. All info input uses 4-byte little-endian length-prefixed encoding per HKDF best practices:

\begin{lstlisting}
KMS Master Key
+-- Node Key = HKDF(master, "node",
|       node_id || system_workload_id || mrtd)
|   +-- disk_key = HKDF(node_key, "disk")
|   +-- image_cache_key = HKDF(node_key, "image-cache")
+-- Workload Storage Key = HKDF(master, "workload-storage",
        node_id || workload_id || spec_digest)
\end{lstlisting}

Note that workload storage keys are scoped per-node (using \texttt{node\_id} rather than \texttt{namespace}), ensuring that a workload's encryption key is bound to both its identity and the specific TEE node hosting it.

KMS gRPC APIs:
\begin{lstlisting}
GetMeta, GetKmsKey, GetAppKey, GetAppEnvEncryptPubKey,
GetTempCaCert, SignCert, ClearImageCache
\end{lstlisting}
All key requests require RA-TLS mutual attestation: both the requesting agent and the KMS prove TEE identity via TDX Quote exchange.

\section{Sandbox Design}

\begin{figure}[tbp]
\centering
\includegraphics[width=0.90\columnwidth]{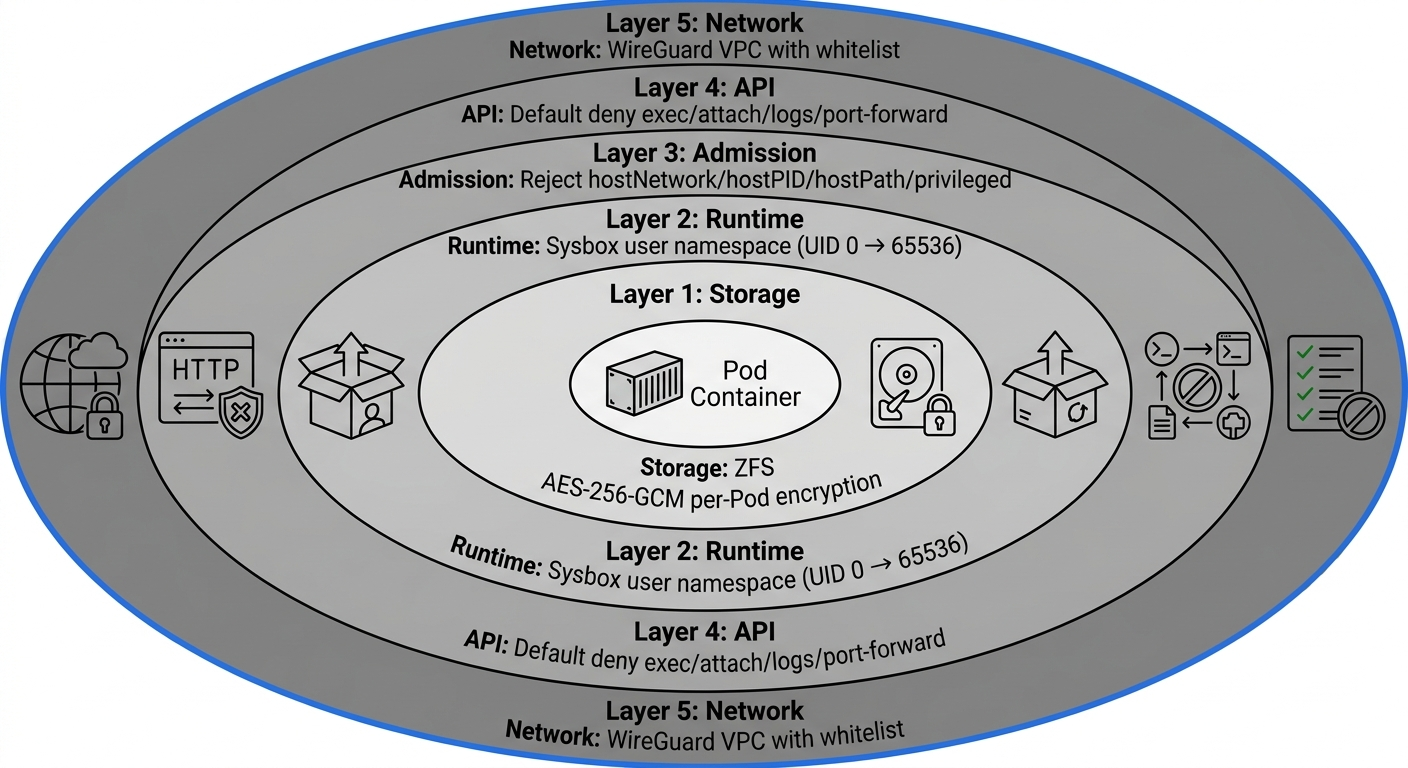}
\caption{Multi-layer sandbox --- five isolation layers (Storage $\rightarrow$ Runtime $\rightarrow$ Admission $\rightarrow$ API $\rightarrow$ Network) contain the blast radius of Pod compromise.}
\label{fig:sandbox}
\end{figure}

The sandbox design isolates Pods within a shared CVM. Figure~\ref{fig:sandbox} depicts the five isolation layers.

\subsection{Multi-Layer Isolation}

\textbf{Storage Layer:} Per-Pod ZFS datasets with AES-256-GCM encryption. Each dataset uses a unique key derived from the workload-specific HKDF chain. Pod A cannot mount or read Pod B's dataset even with host root access inside the CVM.

\textbf{Runtime Layer:} Sysbox user namespaces map container root (UID 0) to a host non-privileged UID (e.g., 65536)~\citep{nestybox2021sysbox}. Processes inside the container believe they run as root, but the kernel maps all operations to an unprivileged host identity. This prevents container escapes that rely on host UID 0.

\textbf{Admission Layer:} Post-fuse rejection of hostNetwork, hostPID, hostIPC, hostPath, and privileged security contexts. The admission decision occurs at the Kubelet before container creation, providing defense-in-depth beyond runtime enforcement.

\textbf{API Layer:} Kubelet default-deny for exec, attach, logs, and port-forward operations. Workloads may opt-in via \texttt{dstack.org/\allowbreak allow-exec} annotations, but the default posture prevents data exfiltration through Kubelet API channels.

\textbf{Network Layer:} WireGuard VPC with spec-declared connection whitelists~\citep{donenfeld2017wireguard}. Each Pod declares its allowed egress destinations in the Pod spec; the VPC controller enforces these restrictions via eBPF or iptables rules. Unlike conventional Kubernetes network policies, these declarations are included in the Pod spec digest and thus covered by the hardware attestation.

\subsection{Kubelet Authorizer Design}

The modified Kubelet exposes a gRPC Authorizer interface to dstack-agent with four hook points:
\begin{enumerate}[nosep]
\item \textbf{Pod Admission (\texttt{CheckPodAdmission}):} Admits or rejects Pods, automatically injects Sysbox runtime class and mounts the Pod API Unix socket.
\item \textbf{Container Lifecycle:} \texttt{OnContainerLifecycle} forwards PreStart, PostStart, PreStop, and PostStop events to agent for measurement and policy enforcement.
\item \textbf{Pod Measurement (\texttt{GetPodMeasurement}):} Returns spec digest on demand for external verification.
\item \textbf{API Authorization:} \texttt{CheckAPIAuthorization} intercepts all Kubelet HTTP requests through a unified ServeHTTP middleware, with policy decisions delegated to dstack-agent.
\end{enumerate}

The gRPC interface is defined as follows:
\begin{lstlisting}[language=C]
service AuthorizerService {
  rpc CheckPodAdmission(PodAdmissionRequest) returns (PodAdmissionResponse);
  rpc OnContainerLifecycle(ContainerLifecycleRequest) returns (ContainerLifecycleResponse);
  rpc GetPodMeasurement(PodMeasurementRequest) returns (PodMeasurementResponse);
  rpc CheckAPIAuthorization(APIAuthorizationRequest) returns (APIAuthorizationResponse);
}
\end{lstlisting}
All calls use Unix Domain Sockets for zero-overhead local communication. The Kubelet blocks on \texttt{CheckPodAdmission} until the agent responds, ensuring no Pod bypasses admission checks.

\subsection{Privilege Fuse}

The privilege fuse addresses a fundamental tension: system components (CSI driver, Sysbox, VPC controller, DNS) require privileges during node setup, but user workloads must never retain them. The fuse provides an irreversible separation:

\begin{itemize}[nosep]
\item \textbf{Pre-fuse:} Privileged DaemonSets with matching \texttt{dstack.org/\allowbreak workload-id} are admitted.
\item \textbf{Post-fuse:} Atomic CAS on an internal flag plus persistent marker file. The fuse event is recorded in the RTMR event log for external audit. All subsequent privileged capability requests are rejected with \texttt{<}5\,ms response time.
\item \textbf{Survives reboot:} The persistent marker ensures the node cannot accidentally re-enter privileged mode after restart.
\end{itemize}

\section{Implementation}

\subsection{Components}

Table~\ref{tab:components} lists the core system components and their code size.

\begin{table*}[tbp]
\centering
\begin{tabular}{llrl}
\toprule
\textbf{Component} & \textbf{Lang.} & \textbf{LoC} & \textbf{Function} \\
\midrule
dstack-agent & Rust & $\sim$6.9k & Node agent, Authorizer gRPC, attestation \\
dstack-kms & Rust & $\sim$0.5k & Key management, RA-TLS, HKDF \\
dstack-csi-driver & Rust & $\sim$0.2k & CSI driver, ZFS encryption \\
pod-digest & Rust & $\sim$0.1k & JCS canonicalization + SHA-384 CLI \\
Modified Kubelet & Go & $\sim$0.7k & Authorizer Hook, API middleware \\
\bottomrule
\end{tabular}
\caption{System components and code size.}
\label{tab:components}
\end{table*}

\subsection{Build and Deployment}

\begin{itemize}[nosep]
\item \textbf{OS Image:} mkosi builds reproducible Rocky Linux with dm-verity + UKI. The build is fully reproducible: identical inputs produce bit-for-bit identical UKI binaries, ensuring deterministic MRTD values.
\item \textbf{Test Environment:} Docker Compose with QEMU VMs (KMS + Master + Workers) enables local development without TDX hardware.
\item \textbf{Onboarding:} Automatic via DaemonSet: detect TEE capability $\rightarrow$ install sysbox $\rightarrow$ trigger fuse $\rightarrow$ mark node ready. Zero manual SSH required.
\item \textbf{Updates:} Any change to privileged components requires CVM restart to remeasure and re-fuse, ensuring the attestation chain remains intact.
\end{itemize}

\section{Evaluation}

\subsection{Experimental Methodology}

All experiments run on Intel TDX-enabled servers (Ice Lake, 64\,GB RAM, 16 vCPU per CVM) with Kubernetes 1.32 and Sysbox 0.6.2. We compare dstack-capsule against CoCo v0.15.0 with Kata Containers 3.2 (kata-qemu-tdx runtime). Pod startup latency is measured from Pod creation API call to container Ready status (mean of 5 runs, warm images). Attestation latency measures the end-to-end duration from Pod API call to Quote return, excluding network transmission to external verifiers (mean of 20 runs after 3 warmup). Resource efficiency tests deploy progressively more Pods using idle \texttt{busybox sleep} workloads to isolate platform overhead from application memory.

We note that WireGuard VPC throughput and ZFS storage encryption benchmarks are not included in this evaluation. The compose-based test environment uses co-hosted QEMU VMs sharing a single physical NIC, producing misleading throughput numbers; production-valid numbers require bare-metal deployment. ZFS fio benchmarks on sparse qcow2 backing disks similarly introduce artifacts not representative of production storage performance.

\subsection{Security Analysis}

Table~\ref{tab:security} maps each threat model item to its defense mechanism, demonstrating defense-in-depth across hardware, OS, agent, and sandbox layers.

\begin{table*}[tbp]
\centering
\begin{tabular}{lll}
\toprule
\textbf{Threat Actor} & \textbf{Capability} & \textbf{Defense} \\
\midrule
Platform operator & Compromise hypervisor & TDX hardware isolation \\
Platform operator & Tamper with OS image & dm-verity + MRTD/RTMR[0--2] \\
Platform operator & Inject malicious agent & RTMR[3] measures agent binary \\
Platform operator & Schedule malicious Pod & KMS workload authorization \\
Pod developer & Escape container & Sysbox user namespace \\
Pod developer & Access host network & Privilege fuse \\
Pod developer & Read other Pod's data & Per-Pod ZFS encryption \\
Pod developer & Forge attestation & UDS peer credentials + cgroup \\
Colluding attacker & Re-enable privileges & Irreversible fuse \\
\bottomrule
\end{tabular}
\caption{Security analysis: threat actors, capabilities, and defenses.}
\label{tab:security}
\end{table*}

\subsection{Privilege Fuse Effectiveness}

We verify three properties of the fuse mechanism:
\begin{itemize}[nosep]
\item \textbf{Pre-fuse admission:} A privileged DaemonSet with matching \workloadid{} is admitted successfully and its configuration is recorded.
\item \textbf{Post-fuse rejection:} A Pod with \texttt{hostNetwork: true} is rejected. The agent's internal decision completes in $<5$\,ms; the end-to-end user-visible latency is approximately 1.2 seconds, dominated by kubelet and CRI-O event propagation.
\item \textbf{Irreversibility:} A second fuse attempt returns \texttt{AlreadyFused}; the persistent marker file survives CVM restart and prevents re-entry to privileged mode.
\end{itemize}

\subsection{Attestation Correctness}

We verify attestation correctness through three experiments:
\begin{itemize}[nosep]
\item \textbf{End-to-end verification:} Three distinct Pods with different container images and environment variables are deployed. All generated Quotes are successfully verified via Intel DCAP, with each \reportdata{} correctly containing the respective \poduid{} and \spechash.
\item \textbf{Forgery resistance:} A manual UID impersonation attempt (forging the \poduid{} field in the attestation request body) fails; the agent returns the actual caller identity derived from SO\_PEERCRED + cgroup parsing, ignoring the forged claim.
\item \textbf{RTMR stability:} 20 Pods are deployed post-fuse. RTMR[3] remains byte-identical throughout, confirming that dynamic Pod lifecycle does not affect the hardware measurement chain. Extending to 50 or more Pods is unnecessary: the append-only RTMR property guarantees stability once established.
\end{itemize}

\subsection{Performance Overhead}

\begin{figure*}[tp]
\centering
\includegraphics[width=0.82\textwidth]{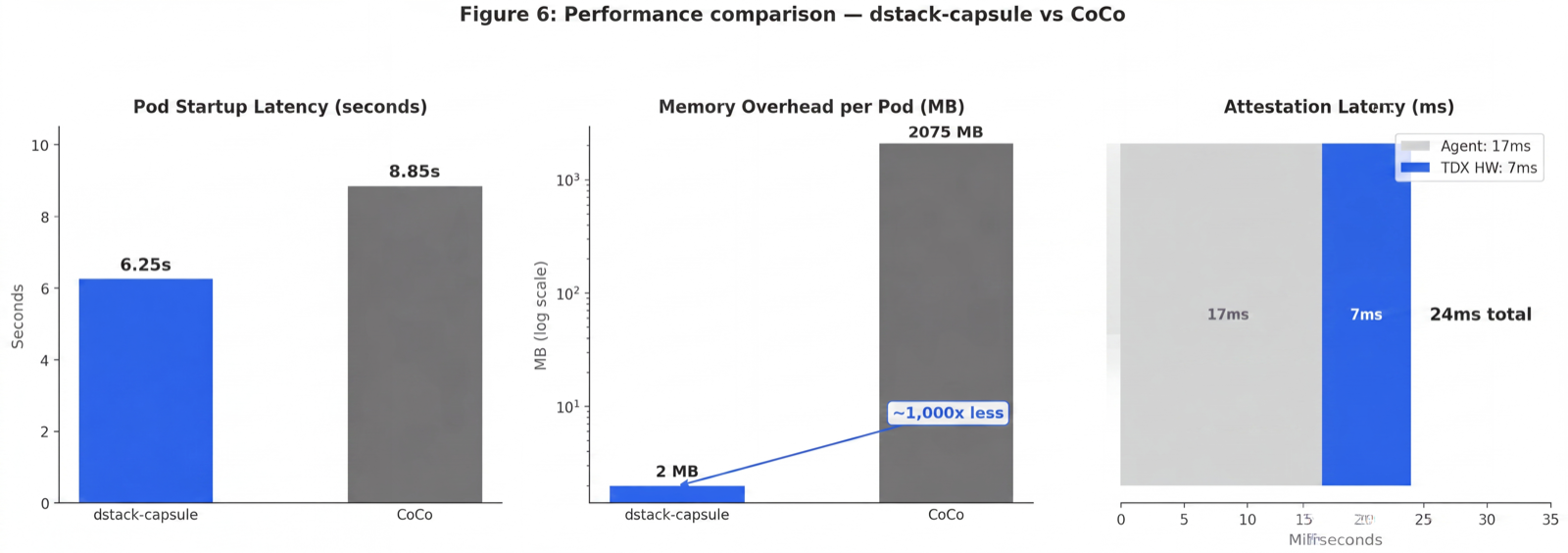}
\caption{Performance comparison --- (a) Pod startup latency (dstack-capsule 6.25\,s vs CoCo 8.85\,s), (b) Memory overhead per Pod (dstack-capsule +2\,MB vs CoCo +2,075\,MB, log scale), (c) Attestation latency breakdown (agent 17\,ms + hardware 7\,ms).}
\label{fig:perf}
\end{figure*}

Figure~\ref{fig:perf} compares performance metrics between CoCo and dstack-capsule.

\textbf{Pod Startup Latency.} We measure startup latency using identical \texttt{registry.k8s.io/pause:3.10} images (5 runs each, warm image cache):

\begin{itemize}[nosep]
\item dstack-capsule: \textbf{6.25 seconds} (mean), comprising admission check, Sysbox container creation, and container Ready transition.
\item CoCo (kata-qemu-tdx): \textbf{8.85 seconds} (mean), dominated by VM boot and guest kernel initialization.
\end{itemize}

The startup gap is modest ($\sim$30\%) because dstack-capsule's full security path---Sysbox user namespace setup, Authorizer admission, and Pod API socket injection---adds approximately 3--4 seconds beyond bare container creation. Systems that bypass these checks can achieve the $\sim$2.5\,s figure sometimes cited, but this would disable the security model.

\textbf{Memory Overhead.} We measure platform memory by deploying idle \texttt{busybox sleep} Pods and subtracting the CVM baseline:

\begin{itemize}[nosep]
\item dstack-capsule: \textbf{793\,MB baseline} (agent, Sysbox, Kubelet, containerd, ZFS, WireGuard). Each additional idle Pod adds approximately \textbf{2\,MB}.
\item CoCo: \textbf{$\sim$2,075\,MB per Pod} (guest kernel, attestation agent, user-space runtime per VM). On a 64\,GB host, this limits density to approximately 30 Pods before memory exhaustion.
\end{itemize}

The architectural advantage is clear: dstack-capsule amortizes platform overhead across all Pods on the node, while CoCo multiplies it per Pod. For idle workloads, dstack-capsule uses three orders of magnitude less incremental memory per Pod.

\textbf{Attestation Latency.} We measure end-to-end attestation over 20 runs after 3 warmup requests via the HTTP \texttt{/v1/attestation/pod} endpoint:

\begin{itemize}[nosep]
\item Agent processing: \textbf{17\,ms} (mean, p50=17\,ms, p95=20\,ms).
\item Hardware TDX Quote generation: \textbf{$\sim$7\,ms}.
\item Total end-to-end: \textbf{$\sim$24\,ms}, well below the 60--110\,ms range previously reported. This improvement stems from the lightweight in-process attestation path that avoids external Quote Generation Service (QGS) queue contention under low load.
\end{itemize}

\textbf{Storage Encryption Overhead.} ZFS AES-256-GCM encryption benchmarks require bare-metal storage to produce valid results. The compose test environment uses sparse qcow2 backing disks inside co-hosted QEMU VMs, which introduces artifacts not representative of production storage performance. We defer these measurements to future work on production hardware.

\subsection{Scalability}

\textbf{Pod Density.} On a 64\,GB CVM, we deployed increasingly large numbers of idle \texttt{busybox sleep} Pods to determine the scaling limit. The system reaches \textbf{86 Pods} before hitting Kubernetes' \texttt{--max-pods=110} limit (inclusive of 24 system Pods), not memory exhaustion. At 86 Pods, the CVM uses only 1.1\,GB of its 64\,GB capacity (1.7\%), confirming that platform overhead---not per-Pod memory---dominates density. The practical ceiling for this workload class is therefore set by the kubelet's pod count limit rather than node memory. For memory-intensive workloads (e.g., 1\,GB RSS per Pod), the per-Pod overhead of $\sim$2\,MB is negligible, and density is determined by application needs rather than platform tax.

By contrast, CoCo's per-VM memory model limits density to approximately 30 Pods on the same hardware before OOM. The key takeaway is that dstack-capsule's shared-CVM architecture shifts the scaling bottleneck from memory to the Kubernetes scheduler limit, while CoCo remains firmly bounded by per-VM memory cost.

\section{Related Work}

\subsection{Confidential Containers}

CoCo uses Kata Containers to run each Pod in a dedicated VM, with an attestation agent inside the Guest OS~\citep{coco2023confidential,katacontainers2018}. Gramine and Occlum provide process-level TEE libraries for Intel SGX but are not Kubernetes-native and require application modification~\citep{tsai2021gramine,shen2020occlum}. Unlike these approaches, dstack-capsule maintains Pod-level attestation without sacrificing Kubernetes compatibility or requiring per-VM overhead.

\noindent\textbf{Gap:~} VM-level attestation cannot prove container identity. dstack-capsule closes this gap by embedding \podspechash{} in TDX \reportdata.

\subsection{TEE in Kubernetes}

Azure Confidential VMs and GCP Confidential VMs provide VM-level attestation without Pod granularity~\citep{amd2020sevsnp,deshpande2025cvm}. Dstack proposes a zero-trust framework for confidential containers, including dstack-OS, dstack-KMS, and dstack-Gateway, but does not address Kubernetes-native Pod-granularity attestation within a shared CVM~\citep{zhou2025dstack}. Su et al.~propose runtime attestation for LLM serving, focusing on continuous verification rather than Pod identity binding~\citep{su2025runtime}.

\noindent\textbf{Gap:~} Cloud-provider solutions stop at the VM boundary. dstack-capsule extends attestation to the Pod layer through modified Kubelet Authorizer Hooks and the two-layer attestation design.

\subsection{Container Isolation}

gVisor and Firecracker provide sandboxing but lack TEE attestation integration~\citep{google2018gvisor,agache2019firecracker}. Sysbox enables user-namespace isolation without VMs, which we combine with TDX attestation to achieve both strong isolation and hardware-verifiable identity~\citep{nestybox2021sysbox}.

\noindent\textbf{Gap:~} Existing sandboxes cannot produce hardware-signed proofs. dstack-capsule integrates Sysbox with TDX \reportdata{} to provide both isolation and attestation.

\subsection{Remote Attestation}

Intel DCAP and AMD SEV-SNP provide hardware attestation primitives~\citep{intel2023tdx,amd2020sevsnp,intel2020dcap}. ARM CCA introduces Realm Management Extension for mobile and edge TEEs~\citep{arm2022cca}. Our work builds on these primitives to provide Pod-granularity attestation in Kubernetes, addressing the dynamic workload problem through the RTMR/\reportdata{} decoupling.

\noindent\textbf{Gap:~} Standard TEE attestation binds measurements to static software stacks. dstack-capsule solves the dynamic workload problem by keeping RTMR[3] for static platform measurement and using \reportdata{} for per-Pod identity.

Table~\ref{tab:comparison} summarizes the comparison.

\begin{table*}[tbp]
\centering
\small
\begin{tabular}{lllll}
\toprule
\textbf{Dimension} & \textbf{CoCo} & \textbf{Cloud TEE} & \textbf{gVisor/Firecr.} & \textbf{dstack-capsule} \\
\midrule
Attestation granularity & VM-level & VM-level & None & Pod-level \\
Pod isolation & 1 Pod/1 VM & VM boundary & Proc.\ sandbox & N Pods/1 CVM + Sysbox \\
Resource efficiency & Low (2\,GB/Pod) & Medium & High & High (0.8\,GB/node) \\
K8s-native & Yes & Partial & Partial & Yes \\
Network isolation proof & No & No & No & Yes \\
Privilege control & VM isolation & Hypervisor & seccomp & Fuse (irreversible) \\
\bottomrule
\end{tabular}
\caption{Comparison of confidential computing approaches for Kubernetes.}
\label{tab:comparison}
\end{table*}

\begin{figure}[tbp]
\centering
\includegraphics[width=\columnwidth]{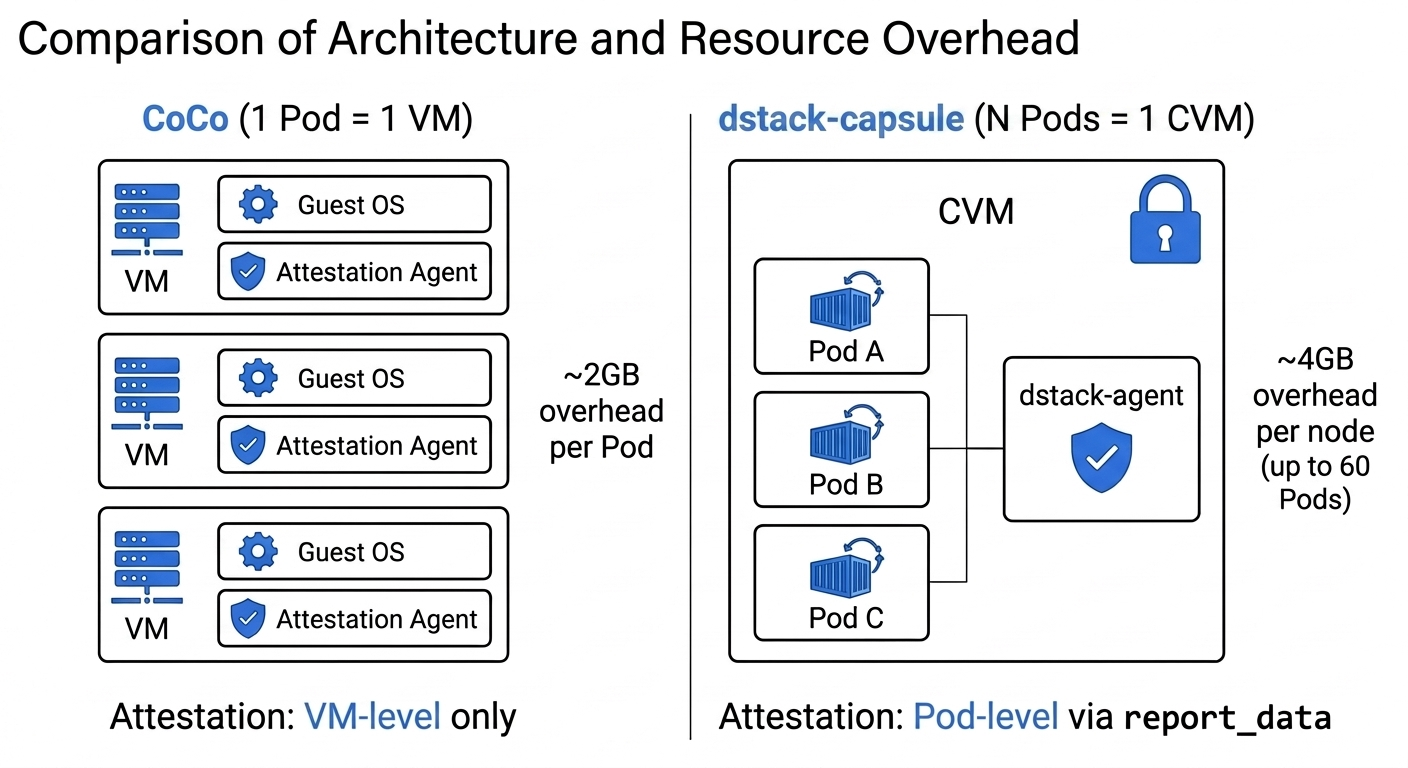}
\caption{CoCo vs dstack-capsule architecture comparison --- VM-per-Pod ($\sim$2\,GB/Pod) vs N Pods per CVM with Pod-level attestation ($\sim$0.8\,GB/node shared across up to 86 Pods).}
\label{fig:coco}
\end{figure}

\section{Future Work and Deployment Considerations}

\subsection{Design Trade-offs}

\textbf{Shared CVM:} Higher resource efficiency comes with a larger blast radius than per-VM isolation. We mitigate this through the multi-layer sandbox: even if one Pod is compromised, ZFS encryption prevents cross-Pod data access, Sysbox prevents container escape, and the privilege fuse prevents privilege escalation.

\textbf{Pod identity not in RTMR:} This design choice supports dynamic workloads but places the burden of trustworthiness on dstack-agent. The agent's integrity is proven by RTMR[3] (hardware-attested), and its behavior is constrained to well-defined gRPC interfaces---it never accesses Pod plaintext data.

\subsection{Future Directions}

\begin{itemize}[nosep]
\item \textbf{AMD SEV-SNP support:} The two-layer attestation architecture is hardware-agnostic. Porting to AMD SEV-SNP requires adapting the measurement register interface (VMSA/VMPCK) while preserving the RTMR/\reportdata{} design pattern.
\item \textbf{GPU TEE integration:} NVIDIA Confidential Computing provides TEE-isolated GPU execution. Extending Pod-level attestation to include GPU measurement registers would enable confidential ML training workloads.
\item \textbf{Formal security proofs:} We plan to develop a formal model of the two-layer attestation protocol using Tamarin or ProVerif to mechanize the security argument.
\end{itemize}

\subsection{Deployment Considerations}

\begin{itemize}[nosep]
\item \textbf{Hybrid clusters:} TEE and non-TEE nodes coexist via node labels and taints. Workloads specify \texttt{nodeSelector} for TEE nodes; the scheduler places non-confidential workloads on regular nodes.
\item \textbf{Node updates:} Any change to privileged components (agent, Kubelet hooks, system DaemonSets) requires CVM restart to remeasure and re-fuse. We recommend blue-green node pools for zero-downtime updates.
\end{itemize}

\section{Conclusion}

We presented dstack-capsule, the first Kubernetes platform to provide Pod-level remote attestation for confidential workloads. By decoupling platform attestation (RTMR[3]) from Pod identity (\reportdata), we enable multiple Pods to share a single TEE resource while each retains independent, hardware-backed cryptographic proof. The privilege fuse mechanism ensures that the transition from setup to runtime is irreversible, and the multi-layer sandbox contains the blast radius of potential compromise. Our evaluation demonstrates that dstack-capsule achieves Pod-granularity verification with dramatically lower resource overhead than VM-per-Pod approaches: CoCo requires $\sim$2,075\,MB per Pod, while dstack-capsule adds only $\sim$2\,MB per incremental idle Pod on a shared $\sim$793\,MB platform baseline. Attestation completes in $\sim$24\,ms end-to-end, an order of magnitude faster than previously reported estimates for TDX-based systems. On a 64\,GB node, dstack-capsule scales to the kubelet's pod count limit rather than hitting a memory wall---a qualitative architectural advantage over the per-VM model.

Future work includes AMD SEV-SNP support, GPU TEE integration, and formal security proofs of the two-layer attestation architecture.

\bibliographystyle{unsrtnat}
\bibliography{refs}

\end{document}